\documentclass[11pt]{article}

\usepackage[T1]{fontenc}
\usepackage{textcomp}
\usepackage[centertags]{amsmath}
\usepackage[mediummath]{nccmath}
\usepackage{amsfonts}
\usepackage{amssymb}
\usepackage[pdftex]{hyperref}
\usepackage{graphicx}
\usepackage[numbers,sort&compress]{natbib}

\usepackage{paperinitial}

\paperinitialization{15mm}{15mm}{15mm}{15mm}{2pt}{10pt}

\DeclareMathOperator{\rot}{rot}

\newcommand{\lan}{\langle}
\newcommand{\ran}{\rangle}

\newcommand{\vf}{\varphi}

\newcommand{\s}{\sigma}

\newcommand{\Ga}{\Gamma}
\newcommand{\de}{\delta}

\newcommand{\spx}{\mathbf{x}}

\newcommand{\spp}{\mathbf{p}}
\newcommand{\spk}{\mathbf{k}}
\newcommand{\spn}{\mathbf{n}}
\newcommand{\spe}{\mathbf{e}}

\newcommand{\spR}{\mathbf{R}}

\newcommand{\R}{\mathbb{R}}

\newcommand{\bs}{\boldsymbol}

\begin{document}
\allowdisplaybreaks[4]
\frenchspacing

\title{{\Large\textbf{Multiplexing signals with\\ twisted photons by a circular arc phased array}}}

\date{}

\author{P.O. Kazinski\thanks{E-mail: \texttt{kpo@phys.tsu.ru}},\;
P.S. Korolev\thanks{E-mail: \texttt{kizorph.d@gmail.com}},\;
G.Yu. Lazarenko\thanks{E-mail: \texttt{laz@phys.tsu.ru}},\;
and V.A. Ryakin\thanks{E-mail: \texttt{vlad.r.a.phys@yandex.ru}}\\[0.5em]
{\normalsize Physics Faculty, Tomsk State University, Tomsk 634050, Russia}
}

\maketitle

\begin{abstract}

The theory of multiplexing electromagnetic signals by means of twisted photons generated by a uniform circular array (UCA) is developed in the case when the receiving antenna represents an array of elements located on a circular arc. The radiating elements are characterized by certain current distributions and are not points, in general. The polarization of created electromagnetic waves is fully taken into account. The notion of discrete twisted photons of the order $N$ is introduced and orthogonality of these modes modulo $N$ is established. Both paraxial and planar discrete twisted photons are considered. The explicit expressions for the signals received are obtained. It is shown that, in the simplest scenario, a $K$ times decrease of the circular arc where the receiving array antenna is placed results in a $K$ times decrease of the number of independent information channels. In the more sophisticated approach, one can restore all $N\gg1$ independent information channels in receiving the signal by an array antenna with $N$ elements located on a circular arc with the central angle $2\pi/K$. However, this problem becomes rapidly ill-conditioned as one increases $K$. The method mitigating this issue is described. The estimates for the corresponding condition numbers are found. The scenario with beam steering, where the radiation produced by the UCA is concentrated near the receiving circular arc array antenna, is also investigated. The orthogonality of the information channels is proved in this case and the corresponding transformation matrix and its condition number are found.

\end{abstract}

\section{Introduction}

Nowadays there are elaborated methods for generating and receiving electromagnetic waves with definite projections of the total angular momentum \cite{Roadmap16,SerboNew,ZWYB20,Noor22,Willner22,JiangWerner22}. Following \cite{SerboNew,MolTerTorTor} we will refer to the modes of the electromagnetic field with definite projection of the total angular momentum onto a certain axis as the twisted photons. The most developed technology for producing and detecting twisted photons in the radio frequency domain is based on the usage of uniform circular arrays (UCAs) \cite{ZWYB20,Noor22,Thide07,Mohammadi10}. This approach allows one to multiplex and to demultiplex readily the modes with distinct projections of angular momentum in addition to the standard methods for increasing the density of information transfer \cite{PSNC14,Bai17,CZLJL18,Zhu20,LZXPW20,YaSaYaLe21,AssAbbFus21,LCMXL21,Zhang22,Murata2022,YaSaLee22,CZLZ23}. It was implemented not only in the radio frequency domain but also in the THz \cite{Khan22,Willner22} and optical \cite{ADKL20} ranges. One of the drawbacks of long-range information transmission with the help of the twisted photons is the conical divergence of the intensity of modes with nonzero projection of the orbital angular momentum at large distances \cite{Li20,Papathanasopoulos2022}. This quite large divergence stems from the fact that the intensity of twisted modes on the propagation axis is zero for nonvanishing projections of the orbital angular momentum. As a result, the receiving UCA antenna should be large in order to accumulate the signal of a considerable intensity. This problem can be mitigated if one receives and demultiplexes the signal with a sufficiently small circular arc array antenna and employ beam steering in order to concentrate the radiation on the receiving antenna \cite{Zheng2015}. In the present paper, we develop the formalism describing such a scenario and propose several methods for transmission of independent signals by means of twisted photons in such a configuration of antennas.

One may distinguish the two main approaches to transmit information with the aid of twisted photons. They are based on paraxial and planar solutions to the Maxwell equations with definite projections of the total angular momentum. In the first case, the intensity of radiation is concentrated in a narrow cone with the axis coinciding with the quantization axis of the projection of total angular momentum. This approach is suitable for a point-to-point information transmission. In the second case, the intensity of radiation is concentrated near the plane orthogonal to the quantization axis and the twisted photons propagate mainly in this plane carrying the energy from the quantization axis to infinity \cite{LCMXL21,AssAbbFus21,LZXPW20,Zhu20}. This method is useful for a point-to-multipoint communication. We consider in the paper the both cases and elaborate the corresponding theory for multiplexing the signals by a UCA and demultiplexing these signals by a circular arc array antenna.

Notice that receiving of the electromagnetic waves with nonzero projections of the orbital angular momentum by a circular arc array antenna was already discussed in the literature \cite{Mohammadi10rs,ZhangMa17,ZhaoZhang19,Chen2020,Zheng2018,Zheng2022,Chen2022}. Nevertheless, the approaches investigated in the present paper are completely different from that proposed in \cite{Mohammadi10rs,ZhangMa17,ZhaoZhang19,Chen2020}. The possible hardware realization of the schemes we describe is based on the use of the respective Butler matrices combining the outgoing and incoming signals in the array antennas. The scenarios we consider are close to the ones described in \cite{Zheng2018,Zheng2022} but in our approach the twisted photons are produced by a UCA and not by a horn antenna. In the papers \cite{Qasem2021,Qasem2022,Klemes2016,Yin2019}, the beam steering by a UCA was considered but a multiplexing scheme was not elaborated. In the paper \cite{Chen2022}, the twisted radiowaves created by a UCA and received by a circular arc array antenna were investigated. However, the finite size and, correspondingly, the radiation patterns of radiating elements of the UCA were not taken into account in this work. Moreover, beam steering and the respective improvement of the scheme for information transmission was not discussed there.

As regards the transmitting and receiving UCAs, such a scheme was realized in \cite{YaSaLee22,YaSaYaLe21,PSNC14}. It should be stressed that, in accordance with this method and as it follows from the theory we develop, the number of UCA elements, $N$, can be small and the whole scheme works even for $N=2$. It is a consequence of exact orthogonality of the modes of the electromagnetic field produced by the UCA even in the case of a finite $N$ \cite{YaSaLee22,Li20,Murata2022}. The different radiated modes are comprised of twisted photons with different projections of the total angular momentum.

The paper is organized as follows. In Sec. \ref{Gen_Formul}, we introduce the notation and provide some general formulas used in the subsequent sections. In Sec. \ref{Indep_Channels}, we start with a short exposition of the complete set of solutions of the free Maxwell equations in a vacuum that describes twisted photons \cite{JaurHac,BKL2,SerboNew}. The explicit expressions for the particular cases of these solutions corresponding to paraxial and planar twisted photons are also discussed. In Secs. \ref{simpl_model}, \ref{Planar_Tw_Phot}, we develop the formalism to describe multiplexing and demultiplexing of signals by means of the discrete analogues of paraxial and planar twisted photons. Section \ref{CAAA} is devoted to demultiplexing of the signal with the aid of a circular arc array antenna, the signal being created by the UCA. We develop the corresponding theory and find, in particular, the estimates for the conditions numbers of the respective signal transformation matrices. In Conclusion, we summarize the results. In Appendix \ref{Discr_Bessel_App}, the properties of the discrete Bessel functions \cite{BCFT16,UriWo20,UriWo21,UriWo21-1,WanSzekAf22,WanAf22} describing the electromagnetic field produced by a UCA are outlined. Appendix \ref{Vandermon_Matrix} is devoted to the main properties of the Vandermonde matrix \cite{BagMitr99,Marvasti21} that appears in processing of the signal received by a circular arc phased array.

We use the system of units such that $\hbar=c=1$. In particular, we do not distinguish the photon momentum and its wave vector. The Einstein summation notation is also assumed. We identify the axes $1$, $2$, and $3$ with the axes $x$, $y$, and $z$.

\section{General formulas}\label{Gen_Formul}

The vector potential in the wave zone in a vacuum has the form \cite{LandLifshCTF.2}
\begin{equation}\label{field_gener}
    A_i(k_0;\spR)=\frac{e^{i|\spk|R}}{R} j_i(k_0,\spk),\qquad\spk:=k_0\spn,\qquad \spn=\spR/R,
\end{equation}
where the vector $\spR$ is directed from the center of the antenna to the observation point, $R=|\spR|$, and
\begin{equation}\label{current_Fourier}
    j_i(k_0,\spk):=\int dtd\spx e^{i(k_0t-\spk\spx)} j_i(t,\spx),
\end{equation}
where $j_i(t,\spx)$ is the current density of the radiating system. If $j_i(t,\spx)$ are absolutely integrable functions with compact supports, then their Fourier transforms \eqref{current_Fourier} are entire analytic functions of $k_\mu$, $\mu=\overline{0,3}$. Since $j_i(t,\spx)$ are real-valued, we have
\begin{equation}\label{real_valuedness}
    j^*_i(k_0,\spk)=j_i(-k^*_0,-\spk^*).
\end{equation}
The radiation is determined only by $j_i(k_0,\spk)$ taken on the photon mass-shell $k_0=|\spk|$. The electric field strength is given by
\begin{equation}
    E_i=ik_0(A_i-n_i (\spn \mathbf{A})).
\end{equation}
We suppose that the Fourier transform of the current density has the form
\begin{equation}\label{current}
    j_i(k_0,\spk)=\sum_{n=0}^{N-1} V_n(k_0) f^n_i(k) e^{-i\spk\spx_n},
\end{equation}
where
\begin{equation}
    f^n_i(k)=(O_n)_{ij}f_j(k_0,\spk_n),\qquad O^T_nO_n=1,
\end{equation}
and $\spk_n=O^{-1}_n\spk$. In other words, the radiating antenna is an array antenna consisting of $N$ identical elements rotated with respect to each other with the orthogonal matrices $O_n$ and shifted with respect to the antenna center by $\spx_n$. The factor $V_n(k_0)\in \mathbb{C}$ describes the amplitude and the phase of the current applied to the element with the number $n$.

We assume that the receiver obtains the following signal
\begin{equation}\label{detector}
    S_c:=\sum_{r=0}^{M-1} W_{cr} \bs\xi^*_{r}\mathbf{E}(k_0;\spR_r)=ik_0\sum_{r=0}^{M-1} W_{cr} \bs\xi^*_{r}\mathbf{A}(k_0;\spR_r),
\end{equation}
where $\bs\xi_{r}$ specifies the vector of polarization registered by the $r$-th element of the receiving antenna, $W_{cr}$ characterizes the sensitivity of the $r$-th element of the receiver and the additional phase shift that this element adds to the receiving signal and sends it to the output with the number $c$. The second equality in \eqref{detector} holds because $(\bs\xi_r\spk_r)=0$.

\section{Independent signal transmission channels}\label{Indep_Channels}

In the present paper, the independence of the information transmission channels is provided by orthogonality of the electromagnetic field modes possessing definite projections of the total angular momentum. Consider the complete set of solutions of the free Maxwell equations,
\begin{equation}\label{Max_eqs}
    (\rot^2_{ij} -k_0^2\de_{ij})A_j(k_0,\spx)=0,
\end{equation}
in the form of twisted modes of the electromagnetic field (the twisted photons), where $A_i(k_0,\spx)$ is the vector potential in the Coulomb gauge. Suppose that the twisted modes $\bs\psi$ possesses a definite helicity (a circular polarization),
\begin{equation}
    \rot\bs\psi=sk_0\bs\psi,
\end{equation}
where $s=\pm1$ is the photon helicity. Introduce the basis constituted by the eigenvectors of the operator of photon spin projection onto the $z$ axis
\begin{equation}\label{spin_basis}
    \spe_\pm:=\spe_1\pm i\spe_2, \qquad\spe_3,
\end{equation}
where $\{\spe_1,\spe_2,\spe_3\}$ is the right-handed orthonormal triple. Every vector can be expanded in terms of the basis \eqref{spin_basis} as
\begin{equation}\label{basis_expans}
    \bs\psi=\frac{1}{2}(\spe_+\psi_- +\spe_-\psi_+)+\spe_3\psi_3, \qquad\psi_\pm=\psi_1\pm i\psi_2.
\end{equation}
Then the orthogonal complete set of solutions $\bs\psi(s,m,k_3,k_\perp)$ to the Maxwell equations \eqref{Max_eqs} can be cast into the form \eqref{basis_expans} with \cite{JaurHac,BKL2,SerboNew}
\begin{equation}\label{twist_phot}
\begin{split}
    \psi_3(m,k_3,k_\perp)&=\frac{1}{\sqrt{RL_z}}\Big(\frac{n_\perp}{2}\Big)^{3/2} j_m(k_\perp x_+,k_\perp x_-)e^{ik_3x_3},\\
    \psi_\s(s,m,k_3,k_\perp)&=i\frac{s-\s n_3}{n_\perp}\psi_3(m+\s,k_3,k_\perp),
\end{split}
\end{equation}
where $\s=\pm1$, $n_\perp:=k_\perp/|\spk|$, $n_3:=k_3/|\spk|$, $m\in \mathbb{Z}$ is the projection of the total angular momentum onto the $z$ axis, and $1/\sqrt{R L_z}$ is the normalization factor. The Bessel functions $j_\nu(p,q)$ are defined in \eqref{Bessel_j}. In particular,
\begin{equation}
    j_m(k_\perp x_+,k_\perp x_-)=J_m(k_\perp|x_+|)e^{im\vf},
\end{equation}
where $\vf=\arg x_+$. Under the action of the rotation $R_{\vf'}$ by an angle of $\vf'$ around the $z$ axis the solutions \eqref{twist_phot} transform as
\begin{equation}
    R_{\vf'}\bs\psi(s,m,k_3,k_\perp)=e^{im\vf'}\bs\psi(s,m,k_3,k_\perp).
\end{equation}
The solutions \eqref{twist_phot} are divergence-free and orthogonal with respect to the standard scalar product,
\begin{equation}
    \lan\bs\psi,\bs\vf\ran:=\int d\spx \psi^*_i(\spx) \vf_i(\spx),
\end{equation}
for the different quantum numbers $s$, $m$, $k_3$, and $k_\perp$.

In the paraxial limit, $n_3 \approx 1$, we have
\begin{equation}
    \boldsymbol{\psi}(s,m,k_3,k_\perp)\approx \frac{is\spe_s}{\sqrt{4RL_z}}\Big(\frac{n_\perp}{2}\Big)^{1/2} j_l(k_\perp x_+,k_\perp x_-) e^{ik_3z},\qquad l:=m-s,
\end{equation}
where $l\in \mathbb{Z}$ is the projection of the orbital angular momentum onto the $z$ axis. Other particular case of the solutions \eqref{twist_phot} is the so-called planar twisted photons \cite{LCMXL21,AssAbbFus21,LZXPW20,Zhu20} that correspond to $n_\perp\approx1$. These planar twisted photons are generated, for example, by a charged particle moving uniformly along a circle \cite{BKL2,BordKN,EppGus23}. In particular, taking a linear combination of the solutions \eqref{twist_phot}, we obtain the third and plus-minus components of the electromagnetic potential
\begin{equation}
\begin{split}
    \frac12\sum_{s=\pm1} \psi_3(s,m,k_3,k_\perp)&\approx\frac{1}{\sqrt{8RL_z}}j_m(k_\perp x_+,k_\perp x_-)e^{ik_3x_3}\approx\frac{1}{\sqrt{8RL_z}}j_m(k_0 x_+,k_0 x_-),\\
    \frac12\sum_{s=\pm1} \psi_\s(s,m,k_3,k_\perp)&\approx-i\s n_3\psi_3(m+\s,k_3,k_\perp)\approx0,
\end{split}
\end{equation}
where $n_\perp\approx1$ and in the second approximate equalities we have put exactly $n_3=0$ and $n_\perp=1$. These solutions describe the electromagnetic waves propagating in the $(x,y)$ plane with linear polarization $\spe_3$ and the projection of the total angular momentum $m$ onto the $z$ axis. We shall show below how to generate the electromagnetic waves described by ``discrete'' analogues of the aforementioned solutions and how to use their orthogonality to have several independent signal transmission channels with the same frequency and polarization.

\subsection{Uniform circular array}\label{simpl_model}

\begin{figure}[tp]
\centering
\includegraphics*[width=0.75\linewidth]{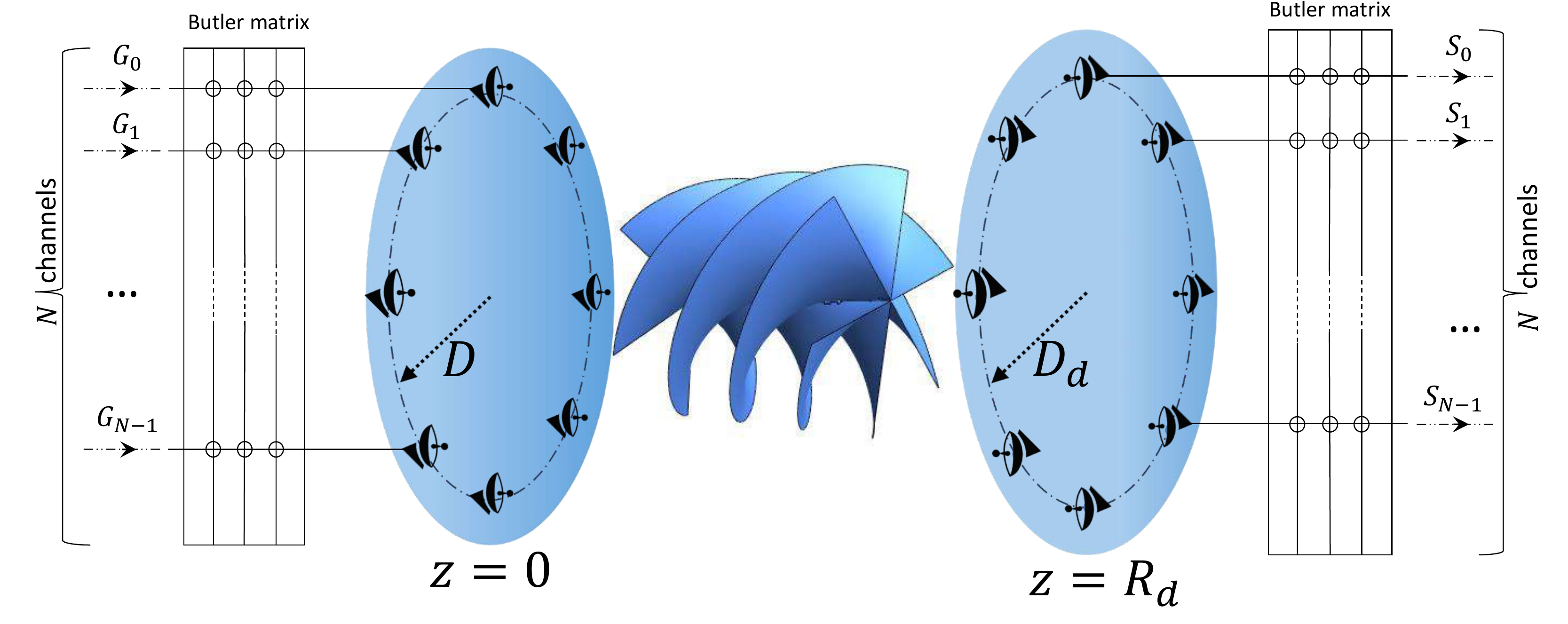}
\caption{{\footnotesize Multiplexing the signals be means of the discrete twisted photons transmitting and receiving by the UCAs.}}
\label{UCAs_plots1}
\end{figure}

Consider the simplest case where the elements of the transmitting array antenna are characterized by the antenna patterns
\begin{equation}\label{source_current}
    f_i(k)=p_i f(k),\qquad \mathbf{p}^*\mathbf{p}=1,
\end{equation}
the complex vector $p_i$ determines the polarization of radiation produced by a single element of the array antenna. The elements are located on the circle of the radius $D$ in the plane $z=0$ (see Fig. \ref{UCAs_plots1}):
\begin{equation}\label{source_array}
    \spx_n=\frac12(\spe_+ x_{n-}+\spe_- x_{n+}),\qquad x_{n+}=De^{i\vf_n},\qquad \vf_n:=2\pi n/N.
\end{equation}
The rotation matrices $O_n=1$. Furthermore, we take
\begin{equation}\label{sourse_signal}
    V_n(k_0)=\sum_{l=0}^{N-1}G_l(k_0) e^{il\vf_n},
\end{equation}
i.e., there are $N$ channels with the signals $G_l$. It is convenient to define $G_{l+kN}=G_l$, $\forall k\in \mathbb{Z}$. Then (see the notation in Appendix \ref{Discr_Bessel_App})
\begin{equation}
\begin{gathered}
    \spk\spx_n=k_\perp D\cos(\phi-\vf_n),\qquad \phi=\arg k_+=\arg(k_1+ik_2),\\
    e^{-i\spk\spx_n}=\sum_{m=-\infty}^\infty(-i)^m J_m(k_\perp D)e^{im(\phi-\vf_n)}\equiv\sum_{m=-\infty}^\infty e^{-im\vf_n} j_m(-ik_+D,ik_-D),
\end{gathered}
\end{equation}
and, consequently,
\begin{equation}
    j_i(k_0,\spk)=p_i f(k)\sum_{n,l=0}^{N-1} G_l\sum_{m=-\infty}^\infty j_m(-ik_+D,ik_-D) e^{i(l-m)\vf_n}.
\end{equation}
Taking into account that
\begin{equation}
    \sum_{n=0}^{N-1} e^{i(l-m)\vf_n}=N\sum_{n=-\infty}^\infty\de_{m,l+Nn}\equiv N\de^N_{ml},
\end{equation}
we come to
\begin{equation}\label{cur_dens_1}
    j_i(k_0,\spk)= p_i f(k)\sum_{l=0}^{N-1}G_l\sum_{n=-\infty}^{\infty}j_{l+Nn}(-ik_+D,ik_-D)= p_i f(k)\sum_{l=0}^{N-1}G_lj_{l}(-ik_+D,ik_-D;N),
\end{equation}
where the discrete Bessel function \eqref{discr_Bess_func} has been introduced.

If $f(k)$ remains unchanged under the transform $\phi\rightarrow\phi+\vf_n$, where $\phi=\arg k_+$, in particular, if $f(k)$ does not depend on $\phi$, then the Fourier transform of the current density can be cast into the form
\begin{equation}\label{cur_dens_tw}
    j_i(k_0,\spk)= \sum_{l=0}^{N-1} G_l\Phi_{i,l}(\phi),
\end{equation}
where, in virtue of the property \eqref{discr_Bess_prop_h}, we have
\begin{equation}\label{eigen_prop}
    \Phi_{i,l}(\phi+\vf_n)=e^{il\vf_n} \Phi_{i,l}(\phi),\qquad \Phi_{i,l+kN}(\phi)=\Phi_{i,l}(\phi),\,\forall k\in \mathbb{Z}.
\end{equation}
This property leads to orthogonality of the modes carrying the signals $G_l$ in the sense that
\begin{equation}
    \int_0^{2\pi}\frac{d\phi}{2\pi} \Phi^*_{i,l'}(\phi)\Phi_{j,l}(\phi)= N\de^N_{l'l} \int_0^{\vf_1}\frac{d\phi}{2\pi} \Phi^*_{i,l'}(\phi)\Phi_{j,l}(\phi).
\end{equation}
As far as the modes entering into \eqref{cur_dens_1} are concerned, their orthogonality property looks as \eqref{orthog_rel}. It is also easy to show that
\begin{equation}\label{inv_Fourier}
    \frac1{N}\sum_{n=0}^{N-1}e^{-il\vf_n}j_i(k_0,\spk)\Big|_{\phi=\phi_0+\vf_n}=G_l\Phi_{i,l}(\phi_0),
\end{equation}
i.e., the different information channels $G_l$ are separated by the inverse discrete Fourier transform. From physical point of view, the property \eqref{eigen_prop} means that the electromagnetic waves created by the $l$-th mode of the current \eqref{cur_dens_tw} consist of the twisted photons with the projections of angular momentum $l+kN$ (see, e.g., formula \eqref{discr_Bess_func}). This guarantees orthogonality of the electromagnetic field modes carrying different signals $G_l$. By analogy with the discrete Fourier transform, one may call such modes discrete twisted photons of the order $N$ \cite{BCFT16,UriWo20,UriWo21,UriWo21-1}. If $f(k)$ is a function of $\phi$ of a general form, then as we shall show the information channels become mixed.

The property \eqref{inv_Fourier} suggests how the simplest receiving antenna should be designed that separates $N$ channels with signals $G_l$. Consider the receiving array antenna consisting of $M=N$ identical elements placed at the circle of the radius $D_d$ on the plane $z=R_d$ with the center at the point $x=y=0$ so that
\begin{equation}\label{detect_simpl}
    \spR_r=\frac{D_d}2\big[\spe_+ e^{-i(\vf_r+\zeta_0)} +\spe_- e^{i(\vf_r+\zeta_0)}\big]+R_d\spe_3,\qquad r=\overline{0,N-1},
\end{equation}
where $\zeta_0$ is some fixed angle. Furthermore, we take the polarization vector of the detector $r$ in the form
\begin{equation}\label{detect_simpl1}
    \xi_{ri}=\frac{p_i-n_{ri}(\spp\spn_r)}{\sqrt{(\spp^*\spp)-(\spp^*\spn_r)(\spp\spn_r)}},\qquad \spn_r=\spR_r/R,\qquad R=\sqrt{D^2_d+R_d^2}.
\end{equation}
It satisfies the transversality condition $(\bs\xi_r\spk_r)=0$. The phase shifts in the elements of the receiving array antenna,
\begin{equation}\label{detect_simpl2}
    W_{cr}=We^{-ic\vf_r},\quad c=\overline{0,N-1},
\end{equation}
can be realized, for example, with the aid of the Butler matrix. Such a scheme was experimentally embodied in \cite{YaSaLee22,YaSaYaLe21,PSNC14}. It follows from \eqref{detect_simpl} that
\begin{equation}
    k_\perp=k_0 n_\perp=k_0 D_d/R,\qquad k_{r+}=k_\perp e^{i(\vf_r+\zeta_0)}.
\end{equation}
Moreover, formula \eqref{detect_simpl1} implies
\begin{equation}
    \bs\xi^*_r\spp=\sqrt{(\spp^*\spp)-(\spp^*\spn_r)(\spp\spn_r)}=\sqrt{1-|\spp\spn_r|^2}.
\end{equation}
In the paraxial limit, $n^2_\perp\ll1$, we have $\spn_r\approx\spe_3$ and
\begin{equation}
    \bs\xi^*_r\spp\approx\sqrt{1-p_3^2},
\end{equation}
i.e., this scalar product does not depend on $r$. In the nonparaxial case, for
\begin{equation}\label{circ_polar}
    \spp=\spe_{s}/\sqrt{2}
\end{equation}
with some fixed $s=\pm1$, we deduce
\begin{equation}
    \bs\xi^*_r\spp=\sqrt{1-\frac{n_\perp^2}{2}}.
\end{equation}
As we see, for any $n_\perp$ the scalar products $(\bs\xi^*_r\spp)$ are independent of $r$ if the elements of the array antenna produce radiation with circular polarization. Henceforth, we assume that $c_p:=(\bs\xi^*_r\spp)$ does not depend on $r$.

It is useful to develop $f(k)$ as a Fourier series
\begin{equation}
    f(k)=\sum_{n=-\infty}^\infty f_n(k_0,k_3,k_\perp)e^{in\phi},\qquad f_n(k_0,k_3,k_\perp)= k_\perp^{|n|}g_n(k_0,k_3,k_\perp^2),
\end{equation}
where $g_n(k_0,k_3,k_\perp^2)$ are some infinitely differentiable functions. Then substituting \eqref{cur_dens_1} into \eqref{field_gener} and \eqref{detector}, we arrive at
\begin{equation}\label{detect_signal}
\begin{split}
    S_c&=ik_0\frac{e^{ik_0 R}}{R}\sum_{r=0}^{N-1}c_p W_{cr}f(\spk_r)\sum_{l=0}^{N-1}G_lj_l(k_\perp D\zeta e^{i\vf_r},k_\perp D\zeta^* e^{-i\vf_r};N)=\\
    &=\tilde{W}N\sum_{l=0}^{N-1}G_l j_l(k_\perp D\zeta,k_\perp D\zeta^*;N) \sum_{k=-\infty}^\infty \de^N_{k,c-l}f_k e^{ik\zeta_0}=\\
    &=\tilde{W}N\sum_{l=0}^{N-1}G_l j_l(k_\perp D\zeta,k_\perp D\zeta^*;N) \sum_{k=-\infty}^\infty f_{c-l+Nk}e^{i(c-l+Nk)\zeta_0},
\end{split}
\end{equation}
where the relation \eqref{sum_rule2} has been used in the second equality. Besides, the notation has been introduced
\begin{equation}
    \tilde{W}=ik_0c_p\frac{e^{ik_0 R}}{R}W,\qquad \zeta:=e^{i(\zeta_0-\pi/2)}.
\end{equation}
If the antenna pattern of every element is invariant under the rotations $\phi\rightarrow\phi+\vf_n$, viz.,
\begin{equation}\label{f_k}
    f_k=\de^N_{k0}f_k,
\end{equation}
then
\begin{equation}\label{s_c_simpl_fin}
    S_c=G_c \tilde{W}Nj_c(k_\perp D\zeta,k_\perp D\zeta^*;N) f(k)\big|_{\phi=\zeta_0}.
\end{equation}
Thus we have $N$ independent signal transmission channels. Expression \eqref{s_c_simpl_fin} implies, in particular, that the narrower the antenna pattern of a separate element $f(k)$, the lesser the dependence of $n_\perp$, where the maximum of the absolute value of \eqref{s_c_simpl_fin} is realized, on the channel number $c$. Of course, there are alternative approaches to decrease the dependence of the divergence of twisted modes of the electromagnetic field on the channel number \cite{YLZGJ19,WKMBE21,Khan22,CZLZ23,CZLJL18,ZWYB20}.

In conclusion of this section, we point out how the above formulas change in the case when the elements of the transmitting antenna are shifted with respect to their positions \eqref{source_array} by some vector $\mathbf{a}_n$. It follows from formula \eqref{current} that this shift results in an additional phase factor
\begin{equation}
    f_i(k)\rightarrow f_i(k)e^{-i\spk \mathbf{a}_n}.
\end{equation}
Therefore, the above formulas remain intact when $k_0 |\spn\mathbf{a}_n|\ll1$ for those values of $k_0$ and $\spn$ where the signal is received. Moreover, the phase differences stemming from small displacements and rotations of the elements of the transmitting and receiving antennas can be compensated by applying the corresponding phase shifts to these emitters and receivers (see for details, e.g., \cite{LGLG17,CXML18,CLWL20,CTLWZ23}).

\subsection{Planar twisted photons}\label{Planar_Tw_Phot}

\begin{figure}[tp]
\centering
\includegraphics*[width=0.55\linewidth]{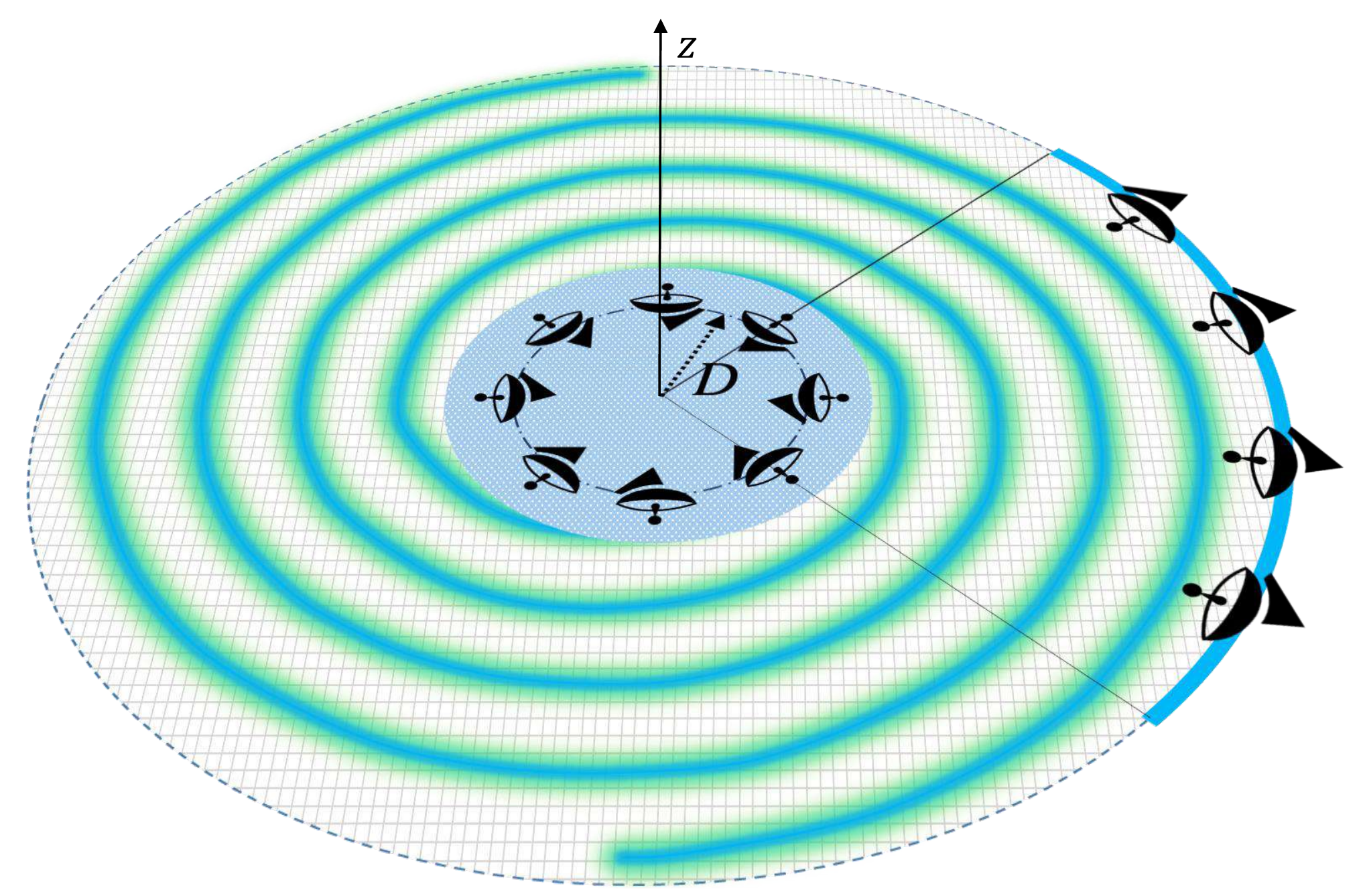}
\caption{{\footnotesize The discrete planar twisted photons radiated by the UCA and received by the circular arc array antenna.}}
\label{UCAs_plots2}
\end{figure}

As is known \cite{LCMXL21,AssAbbFus21,LZXPW20,Zhu20}, the planar twisted photons and their discrete analogues can be produced by the same method as described in Sec. \ref{simpl_model} (see Fig. \ref{UCAs_plots2}). Let us employ the general formulas \eqref{current} and \eqref{detector} with the antenna patterns of the elements \eqref{source_current}. These elements are supposed to be placed at the points \eqref{source_array} and to be fed by the currents \eqref{sourse_signal}. The matrix $O_n$ realizes the rotation by an angle of $\vf_n$ in the plane $(x,y)$ around the $z$ axis so that $\spk_n:=O^{-1}_n\spk$ possesses the components
\begin{equation}
    k_{3,n}=k_3,\qquad k_{+,n}=e^{-i\vf_n} k_+.
\end{equation}
As for the polarization vector of radiation created by the element, $p_i$, we suppose that $\spp=\spe_3$ and, consequently, $O_n\spp=\spp$. Then
\begin{equation}
\begin{split}
    j_i(k)&=p_i\sum_{n,l=0}^{N-1}G_le^{il\vf_n}\sum_{k=-\infty}^\infty f_k e^{ik(\phi-\vf_n)}\sum_{m=-\infty}^\infty j_m(-ik_\perp De^{i\phi},ik_\perp De^{-i\phi})e^{-im\vf_n}=\\
    &=p_iN\sum_{l=0}^{N-1} G_l \sum_{k=-\infty}^\infty j_{l-k}(-ik_+ D,ik_- D;N)f_ke^{ik\phi}.
\end{split}
\end{equation}
Due to the property \eqref{discr_Bess_prop_h}, the functions
\begin{equation}
    \Phi_l(\phi):=N\sum_{k=-\infty}^\infty j_{l-k}(-ik_+ D,ik_- D;N)f_ke^{ik\phi}
\end{equation}
are the eigenfunctions of the discrete rotation operator and obey the relations \eqref{eigen_prop}. As a result, by the same reasoning as was given in Sec. \ref{simpl_model}, we have $N$ independent information transmission channels carried by the discrete planar twisted photons of order $N$.

In order to receive and separate these signal channels, we will act as in Sec. \ref{simpl_model}. We assume that the receiving antenna is a circular array of elements located at the points
\begin{equation}\label{detect_planar}
    \spR_r=\frac{D_d}2\big[\spe_+ e^{-i(\vf_r+\zeta_0)} +\spe_- e^{i(\vf_r+\zeta_0)}\big],\qquad r=\overline{0,N-1},
\end{equation}
and, consequently, $\spn_r=\spR_r/R$, where $R=D_d$. The polarization vectors of the receiving antenna elements are taken in the form $\bs\xi_r=\spp=\spe_3$. Evidently, the transversality condition $(\bs\xi_r\spk_r)=0$ is fulfilled in this case. The phase shifts, $W_{cr}$, giving rise to the channel splitting are set to be equal to \eqref{detect_simpl2}. In that case,
\begin{equation}
    S_c=G_c\tilde{W} N\sum_{k=-\infty}^\infty j_{l-k}(k_0 D\zeta,k_0 D\zeta^*;N)f_ke^{ik\zeta_0},
\end{equation}
where $c_p=1$ in the definition of $\tilde{W}$. As expected, the information channels $G_l$ are separated. Notice that the above formulas are also approximately valid in the case when the locations \eqref{detect_planar} of the elements of the receiving array antenna are the mirror images (for example, in the horn antenna) of their actual locations.

\section{Receiving the signal by a circular arc array antenna}\label{CAAA}

As we have already seen in Sec. \ref{Indep_Channels}, a circular array antenna made of $N$ elements can transmit $N$ independent information channels differing by the projections of angular momentum of photons $l\mod N$. Under certain conditions, this information can be taken by the array antenna consisting of $N$ elements placed on a circular arc concentric with the circular array of transmitting antenna. Such a configuration of the receiving antenna allows one to reduce its size. Furthermore, in the case of information transmission by means of the planar twisted photons in the plane $(x,y)$, as it was described in Sec. \ref{Planar_Tw_Phot}, this method is, in fact, inevitable for a sufficiently large distance from the radiator to the receiver.

Let $M$ elements of the receiving array antenna be located at the points belonging to a circular arc concentric with the circular array of transmitting antenna so that
\begin{equation}
    k_{r+}=k_\perp e^{i\psi_r+i\zeta_0},\qquad\psi_r:=2\pi r/(KM),
\end{equation}
where $K$ specifies the part of the circle where the elements of the receiving antenna are sited, viz., they are placed on the circular arc with the central angle $2\pi/K$ (see Fig. \ref{UCAs_plots2}). Consider the simplest case when $K\in \mathbb{N}$. As in the previous section, we suppose that the elements of the transmitting array antenna are fed by the current \eqref{sourse_signal}, the Fourier transform of the current density has the form \eqref{cur_dens_tw}, where $\Phi_{i,l}(\phi)=p_i\Phi_l(\phi)$, and the scalar products, $\bs\xi_r\spp=c_p$, do not depend on $r$.

If $N=KM$, then the relations \eqref{eigen_prop} imply that the receiving signal can be written as
\begin{equation}\label{S_c_arc_s}
    S_c=ik_0\frac{e^{ik_0 R}}{R}c_p\sum_{r=0}^{M-1} W_{cr}\sum_{l=0}^{N-1} G_l\Phi_l(\psi_r+\zeta_0)=ik_0\frac{e^{ik_0 R}}{R}c_p\sum_{r=0}^{M-1} W_{cr}\sum_{l=0}^{N-1} G_l e^{il\psi_r}\Phi_l(\zeta_0).
\end{equation}
It is clear that the information channels $G_l$ are separated provided the signal transformation matrix, $W_{cr}$, is proportional to the inverse to the matrix $e^{il\psi_r}$. In the general case, this matrix is rectangular. To make this matrix invertible, we assume that only those $G_l$ in the current \eqref{sourse_signal} are different from zero that have the number $l$ divisible by $K$ \cite{Chen2022}. Then the receiving signal is given by
\begin{equation}
    S_c=ik_0\frac{e^{ik_0 R}}{R}c_p\sum_{r,l=0}^{M-1} W_{cr} G_{Kl} e^{iKl\psi_r}\Phi_{Kl}(\zeta_0).
\end{equation}
Taking the transformation matrix of the receiving signal in the form of the inverse discrete Fourier transform
\begin{equation}
    W_{cr}=We^{-icK\psi_r},
\end{equation}
we arrive at
\begin{equation}
    S_c=G_{Kc}M\tilde{W}\Phi_{Kc}(\zeta_0),
\end{equation}
i.e., the information channels $G_{Kc}$ are separated. Thus we conclude that, in the simplest case we have just considered, a $K$ times decrease of the circular arc where the receiving array antenna is placed results in a $K$ times decrease of the number of independent information channels: $M=N/K$.

If $N$ is not a multiple of $KM$, then one can consider the case when $N\gg1$ and so
\begin{equation}\label{eigen_prop_cont}
    \Phi_l(\phi+\psi)=e^{il\psi}\Phi_l(\phi),\quad\forall\psi\in \mathbb{R},
\end{equation}
with good accuracy for a certain range of values of $l$. For example, as it is discussed in Appendix \ref{Discr_Bessel_App}, for $N\gg\max(1,|p|,|q|)$ and $|l|<N/2$, the approximate equality takes place
\begin{equation}\label{discr_Bess_appr}
    j_l(p,q;N)\approx j_l(p,q).
\end{equation}
In the case considered in Sec. \ref{simpl_model}, the variables $|p|=|q|=k_\perp D$, whereas for the model given in Sec. \ref{Planar_Tw_Phot}, we have $|p|=|q|=k_0 D$. The functions on the right-hand side of \eqref{discr_Bess_appr} obey the property \eqref{eigen_prop_cont}. Hence, the modes in the expansion \eqref{cur_dens_tw} also comply with this property. Then the receiving signal is written as
\begin{equation}
    S_c\approx ik_0\frac{e^{ik_0 R}}{R}c_p\sum_{r=0}^{M-1} W_{cr}\sum_{l=-[(N-1)/2]}^{[N/2]} e^{il\psi_r} G_l \Phi_l(\zeta_0).
\end{equation}
where we have renumbered the input signals $G_l$ such that $l$ runs a symmetric interval with respect to $l=0$ for odd $N$. In order to obtain the independent channels $G_l$, one needs to take the signal transformation matrix, $W_{cr}$, proportional to the inverse to the matrix
\begin{equation}\label{M_matr}
    H_{lr}:=e^{2\pi ilr/(KM)}=\cos[2\pi lr/(KM)]+i\sin[2\pi lr/(KM)]=:H_{1lr}+iH_{2lr},
\end{equation}
where $l=\overline{-[(N-1)/2],[N/2]}$ and $r=\overline{0,M-1}$. The matrix $H$ is expressed through the Vandermonde matrix \eqref{Vanderm_defn} by formula \eqref{M_V_rel}, where $x=2\pi/(KM)$. For $K\geqslant1$ and $M=N$ this matrix is invertible and (see the notation in Appendix \ref{Vandermon_Matrix})
\begin{equation}\label{H_inv}
    H^{-1}=U^{-1}V^{-1}.
\end{equation}
The explicit expression for the inverse Vandemonde matrix, $V^{-1}$, is presented in formula \eqref{V-1_expl}. Notice the for $N$ odd, the matrix $H$ is normal, i.e., $[H,H^\dag]=0$, and
\begin{equation}
    H_1H_2=H_2H_1=0.
\end{equation}
In this case the matrix $H$ possesses $(N+1)/2$ real and $(N-1)/2$ purely imaginary eigenvalues.

Setting $M=N$ and choosing
\begin{equation}
    W_{cr}=WH^{-1}_{cr},\qquad c=\overline{-[(N-1)/2],[N/2]},
\end{equation}
we come to
\begin{equation}
    S_c=G_c \tilde{W}\Phi_c(\zeta_0).
\end{equation}
To put it another way, when $N$ is odd and the condition \eqref{eigen_prop_cont} holds, which is valid for the models at issue when the approximate equality \eqref{discr_Bess_appr} is satisfied, we have $N$ independent information transmission channels. For $N$ odd, the number $c$ belongs to an interval symmetric with respect to zero. In this case, the number of involved signal channels per the number of elements $N$ is maximal. This follows from the approximate equality \eqref{discr_Bess_appr}, where it is assumed that $|l|<N/2$, and from the fact that the maximum of radiation intensity is realized at a given angle $\arcsin n_\perp$ simultaneously for both signs of the channel number $c$.

Unfortunately, for sufficiently large $K$ and $N$ the matrix $H_{lr}$ is poorly conditioned. For $N\gg1$ and $x=2\pi/(KN)\ll1/N$, the condition number \eqref{cond_num} becomes
\begin{equation}\label{cond_num_1}
\begin{gathered}
    \kappa\approx\frac{N\Ga(2N-1)}{\Ga^3(N)}\Big(\frac{KN}{2\pi}\Big)^{N-1} \approx\frac{e}{\pi\sqrt{2}}\Big(\frac{2eK}{\pi}\Big)^{N-1}\approx 0.612(1.73K)^{N-1},\\
    \lg\kappa\approx(N-1)(\lg K+0.238)-0.213.
\end{gathered}
\end{equation}
Recall that the condition number characterizes the relative accuracy with which the elements of the signal transformation matrix and the signal $G_c \Phi_c$ should be defined for the relative error of the received signal $S_c$ to be small. Therefore, in receiving the signal from a circular arc array antenna by the method we have just considered, the number $N$ has to be not very large. On the other hand, $N$ must not be small so that the approximate equality \eqref{discr_Bess_appr} holds. The optimal values of $N$ seem to be $\{5,7,9\}$.

One can decrease the condition number by transmitting such a signal that
\begin{equation}
    G_l=
    \left\{
      \begin{array}{ll}
        G_l, & \hbox{$l= kK', \,k\in \mathbb{Z}$;} \\
        0, & \hbox{otherwise.}
      \end{array}
    \right.
\end{equation}
The number of independent nonvanishing signal channels $G_l$ is assumed to be equal to $M\leqslant N$. The signal transformation matrix $W_{cr}$ has the dimensions $M\times M$ and is proportional to \eqref{H_inv} with $x=2\pi K'/(KM)$. Its condition number for $x\ll1/M$ reads
\begin{equation}\label{cond_num_2}
    \kappa\approx\frac{M\Ga(2M-1)}{\Ga^3(M)}\Big(\frac{KM}{2\pi K'}\Big)^{M-1},\qquad \lg\kappa\approx(M-1)\Big(\lg \frac{K}{K'}+0.238\Big)-0.213.
\end{equation}
For $x\approx 2\pi/M$, the condition number of the matrix $W_{cr}$ is close to unity. Moreover, in order to diminish the condition number, one can employ the array antenna whose elements are placed on the circular arc at the angles $\psi_r$ specified by the zeros of the Chebyshev polynomial of the first kind.

There is also a fourth way to gain information from a circular arc phased array. This means was discussed in \cite{Zheng2018,Zheng2022} for planar twisted photons generated by horn antennas. We consider the general case where the twisted photons can be paraxial or planar. Besides, the twisted photons are created by an UCA in our approach. As in the simplest scenario, consider the case $N=KM$, $K\in \mathbb{N}$, and take
\begin{equation}\label{applied_signals}
    G_l=g_l e^{-il\chi_0}\Phi_l^{-1}(\zeta_0),\qquad g_{Ks_1+s_2}=g_{Ks_1},\quad s_1=\overline{0,M-1},\;s_2=\overline{0,K-1},
\end{equation}
in formula \eqref{S_c_arc_s} for the received signal. Now information is transmitted though $M$ information channels $g_{Ks_1}$. As we shall see, the choice of the applied signals of the form \eqref{applied_signals} results in beaming of radiated electromagnetic waves near the direction with the azimuth angle $\chi_0+\zeta_0$. Substituting \eqref{applied_signals} into \eqref{S_c_arc_s}, it is easy to obtain that
\begin{equation}
    S_c=ik_0\frac{e^{ik_0 R}}{R}c_p\sum_{r,l=0}^{M-1} W_{cr}\sum_{s_1=0}^{M-1}g_{Ks_1}e^{is_1K(\psi_r-\chi_0)}t_r(\chi_0),
\end{equation}
where
\begin{equation}
    t_r(\chi_0):=\frac{1-e^{iK(\psi_r-\chi_0)}}{1-e^{i(\psi_r-\chi_0)}}.
\end{equation}
It is clear that on taking
\begin{equation}
    W_{cr}=We^{-icK\psi_r}t^{-1}_r(\chi_0),
\end{equation}
the information channels are decoupled
\begin{equation}
    S_c=g_{Kc}M\tilde{W}e^{-icK\chi_0}.
\end{equation}
The condition number of $W_{cr}$ is
\begin{equation}
    \kappa=\frac{\max_r|t_r(\chi_0)|}{\min_r|t_r(\chi_0)|}.
\end{equation}
If one puts $\chi_0=\pi/K$, then $\max_r|t_r(\chi_0)|\approx K$ and
\begin{equation}
    \min_r|t_r(\chi_0)|\approx \frac{1}{\sin[\pi/(2K)]}\approx\frac{2K}{\pi},
\end{equation}
where it has been assumed $K\gg1$. Hence, the condition number $\kappa\approx \pi/2$.

In order to show that the radiation is concentrated near the direction with the azimuth angle $\chi_0+\zeta_0$, we assume for simplicity that $N\gg1$ and consider the cases of paraxial and planar twisted photons discussed in Secs. \ref{simpl_model}, \ref{Planar_Tw_Phot}. Then, as follows from \eqref{eigen_prop}, \eqref{eigen_prop_cont},
\begin{equation}
    \Phi_{i,l}(\phi)/\Phi_{l}(\zeta_0)=v_ie^{il(\phi-\zeta_0)},
\end{equation}
where $v_i$ is some vector independent of $\phi$. Therefore,
\begin{equation}
    j_i(k_0,\spk)=v_i\sum_{l=0}^{N-1}g_l e^{-il\chi_0}e^{il(\phi-\zeta_0)}= v_i \frac{1-e^{iK(\phi-\chi_0-\zeta_0)}}{1-e^{i(\phi-\chi_0-\zeta_0)}}\sum_{s_1=0}^{M-1} g_{Ks_1}e^{iKs_1(\phi-\chi_0-\zeta_0)}.
\end{equation}
It is clear that for large $K$ the maximum of the modulus of this expression is at $\phi=\chi_0+\zeta_0$. We see that the amplitude of radiation received by the elements of the circular arc array antenna is approximately $K$ times greater than in the first scenario described above and so the intensity is $K^2$ times larger. It is also evident that by superimposing several signals of such a type the UCA antenna produces several beams of radiation whose intersection can be made negligible. This allows one to transmit the independent signals to the different receiving antennas.

\section{Conclusion}

Let us sum up the results. We have developed the theory of multiplexing signals by means of the discrete twisted photons produced by a UCA antenna. The independence of different signal channels ensues from exact orthogonality of discrete twisted modes of the electromagnetic field. In describing such a scenario, we have considered both paraxial and planar discrete twisted photons and obtained the explicit expressions for the receiving signals. These expressions involve the discrete Bessel functions \cite{BCFT16,UriWo20,UriWo21,UriWo21-1} whose basic properties are outlined in Appendix \ref{Discr_Bessel_App}. Several approaches for demultiplexing of the signals received by the array antenna with elements placed on a circular arc have been investigated.

We have studied the four cases. In the first case we assume that the number of elements of the transmitting array antenna, $N$, is a multiple of $K$, where $2\pi/K$ is the central angle of the circular arc of the receiving array antenna, and $K$ is a natural number. In this case, the UCA radiator is tuned to transmit only those signals $G_l$ that have the channel numbers $l$ divisible by $K$ \cite{Chen2022}. Then these signals can be safely restored by the receiving array antenna with $N/K$ elements. In the second case when $N\gg1$, the more elaborated method can be used that allows one to restore all $N$ independent information channels for arbitrary $K\geqslant1$ but the problem becomes rapidly ill-conditioned with increasing $K$. The estimate for the respective condition number is given in \eqref{cond_num_1}. This issue can be alleviated if one thins the independent information channels in a way similar to that has been described above. This is the third case we have considered and such an approach, of course, diminishes the number of independent transmitting signals. The estimate for the condition number in this case is presented in \eqref{cond_num_2}. In order to find such estimates, we have obtained in \eqref{sing_num} the asymptotics of the singular values of the Vandermonde matrix which seem to be unknown before. The fourth scenario is the improvement of the first case. The configurations of antennas are the same as in the first case but the signal applied to the transmitting antenna is such that the radiation is mainly concentrated on the arc where it is received by the circular arc array antenna \cite{Zheng2018,Zheng2022}. It has been shown in this case that the number of independent information channels is the same as in the first scenario but the intensity of the received signal is $K^2$ times greater than in the first case. The information channels are exactly orthogonal even for a finite $N$ and the corresponding transformation matrix has the condition number of order unity.

\paragraph{Acknowledgments.}

This study was supported by the Tomsk State University Development Programme (Priority-2030).

\appendix
\section{Discrete Bessel functions}\label{Discr_Bessel_App}

In describing the signal producing and receiving by the array antenna, there appear the functions
\begin{equation}\label{discr_Bess_func}
    j_l(p,q;N):=\sum_{n=-\infty}^\infty j_{l+Nn}(p,q),\qquad N\in \mathbb{N},\quad l\in \mathbb{Z},\quad p,q\in \mathbb{C},
\end{equation}
where \cite{BKL2}
\begin{equation}\label{Bessel_j}
    j_\nu(p,q):=\frac{p^{\nu/2}}{q^{\nu/2}}J_\nu(p^{1/2}q^{1/2}),\qquad\nu\in \mathbb{C},
\end{equation}
and $J_\nu(z)$ are the Bessel functions of the first kind. Some properties of the functions \eqref{discr_Bess_func} were considered in \cite{BCFT16,UriWo20,UriWo21,UriWo21-1,WanSzekAf22,WanAf22} where they are called the discrete Bessel functions. Notice that the functions \eqref{discr_Bess_func} are not defined for $l\in \mathbb{C}$, because in this case the corresponding series are not converging in an ordinary sense. It is clear that
\begin{equation}
    \lim_{N\rightarrow\infty}j_l(p,q;N)=j_l(p,q).
\end{equation}
The approximate equality,
\begin{equation}
    j_l(p,q;N)\approx j_l(p,q),
\end{equation}
holds for $N\gg\max(1,|l|,|p|,|q|)$. If $l\geqslant0$ is less than $N$ and $N\gg\max(1,|p|,|q|)$, then the approximate equalities,
\begin{equation}
    j_l(p,q;N)\approx j_l(p,q),\; \text{for $l<N/2$};\qquad j_l(p,q;N)\approx j_{l-N}(p,q),\;\text{for $l\geqslant N/2$},
\end{equation}
are valid.

The discrete Bessel functions \eqref{discr_Bess_func} possess the properties
\begin{subequations}
\begin{align}
    j_{l+N}(p,q;N)&=j_l(p,q;N),\label{discr_Bess_prop_a}\\
    j_{l}(p,q;N)&=j_l(p,q;2N)+j_{l+N}(p,q;2N),\\
    j_{l}(p,q;N)&=\sum_{n=0}^{M-1}j_{l+Nn}(p,q;NM),\qquad M\in \mathbb{N},\\
    j^*_{l}(p,q;N)&=j_l(p^*,q^*;N),\\
    2\frac{\partial}{\partial p}j_l(p,q;N)&=j_{l-1}(p,q;N),\\
    2\frac{\partial}{\partial q}j_l(p,q;N)&=-j_{l+1}(p,q;N),\\
    j_l(p,q;2N)&=(-1)^lj_{-l}(q,p;2N),\label{discr_Bess_prop_g}\\
    j_l(pe^{i\vf_n},qe^{-i\vf_n};N)&=e^{il\vf_n}j_l(p,q;N),\qquad\vf_n=2\pi n/N,\label{discr_Bess_prop_h}\\
    j_l(0,0;N)&=\sum_{n=-\infty}^\infty \de_{l,Nn}=\de^N_{l0}.
\end{align}
\end{subequations}
Furthermore, employing the addition theorem for $j_\nu(p,q)$ (see (A6) of \cite{BKL2}), we derive the addition theorem for the discrete Bessel functions \cite{UriWo20,UriWo21,UriWo21-1}
\begin{equation}
\begin{split}
    &\sum_{n=-\infty}^{\infty}j_{l-n}(x_+,x_-;N)j_{n}(y_+,y_-)=\sum_{n=-\infty}^{\infty}j_{l-n}(x_+,x_-)j_{n}(y_+,y_-;N)=\\
    &=\sum_{n=0}^{N-1}j_{l-n}(x_+,x_-;N)j_{n}(y_+,y_-;N)=j_{l}(x_++y_+,x_-+y_-;N),
\end{split}
\end{equation}
where $x_\pm=x_1\pm i x_2$, $y_\pm=y_1\pm iy_2$, and $x_{1,2}, y_{1,2}\in \R$.

\begin{figure}[tp]
\centering
\includegraphics*[width=0.97\linewidth]{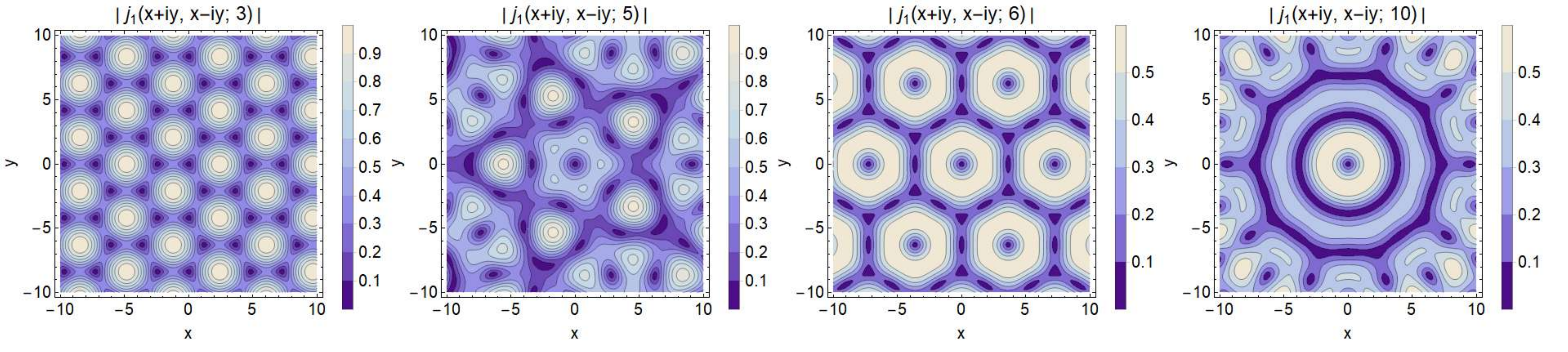}\\
\includegraphics*[width=0.98\linewidth]{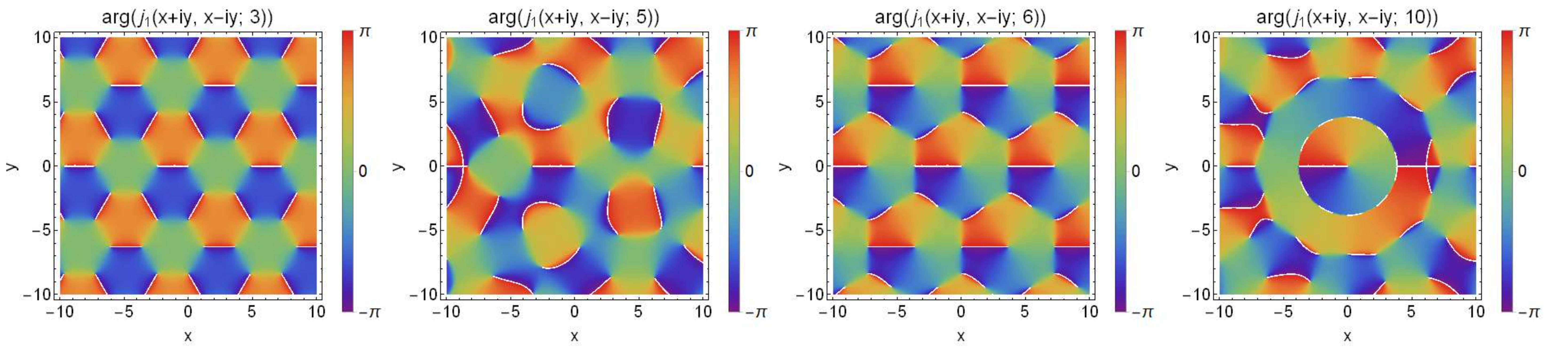}
\caption{{\footnotesize The moduli and arguments of several discrete Bessel functions.}}
\label{discrete_Bessel_plots}
\end{figure}

There are the relations \cite{BKL2,Wats.6}
\begin{equation}\label{gener_func}
    j_l(p,q)=\int_{|t|=1}\frac{dt}{2\pi i}t^{-l-1}e^{(pt-q/t)/2},\qquad\sum_{l=-\infty}^\infty j_l(p,q)t^l=e^{(pt-q/t)/2}.
\end{equation}
Using the first equality or the second one, it is easy to prove that
\begin{equation}
    j_l(p,q;N)=\frac1{N}\sum_{n=0}^{N-1}e^{-il\vf_n}\exp\Big[(pe^{i\vf_n}-q/e^{i\vf_n})/2\Big],
\end{equation}
i.e., $j_l(p,q;N)$ are given by a discrete analogue of the first relation in \eqref{gener_func} and are expressed through elementary functions. The second relation in \eqref{gener_func} implies the sum rule
\begin{equation}
    \sum_{l=0}^{N-1}j_{l}(p,q;N)=e^{(p-q)/2}.
\end{equation}
It follows from \eqref{discr_Bess_prop_h} that
\begin{equation}\label{sum_rule2}
    \frac1{N}\sum_{n=0}^{N-1} e^{-ik\vf_n}j_l(pe^{i\vf_n},q e^{-i\vf_n};N)=\de^N_{kl}j_l(p,q;N).
\end{equation}
In particular, the discrete Bessel functions are orthogonal in the following sense
\begin{equation}\label{orthog_rel}
    \frac1{N}\sum_{n=0}^{N-1} j^*_l(pe^{i\vf_n},q e^{-i\vf_n};N)j_{l'}(p'e^{i\vf_n},q' e^{-i\vf_n};N)=\de^N_{ll'} j^*_l(p,q;N)j_{l}(p',q';N).
\end{equation}
Using the definition \eqref{discr_Bess_func}, it is not hard to show that
\begin{equation}
    \int_{-\pi}^\pi \frac{d\vf}{2\pi}e^{-ik\vf}j_l(pe^{i\vf},qe^{-i\vf};N)=\de^N_{lk}j_k(p,q).
\end{equation}
The functions $|j_l(re^{i\vf},re^{-i\vf};N)|$, $r>0$, have extrema at the points
\begin{equation}
    \vf=\pi/N+\vf_n,\qquad \vf=\vf_n,\qquad n\in \mathbb{Z},
\end{equation}
as functions of $\vf$.

The explicit expressions for some small $N$:
\begin{equation}\label{discr_Bess_func_expl}
\begin{split}
    j_0(x+iy,x-iy;1)&=e^{iy},\\
    j_0(x+iy,x-iy;2)&=\cos y,\\
    j_1(x+iy,x-iy;2)&=i\sin y,\\
    j_0(x+iy,x-iy;3)&=\frac{e^{-iy/2}}3 (e^{3iy/2}+2\cos\frac{\sqrt{3}x}{2}),\\
    j_1(x+iy,x-iy;3)&=\frac{e^{-iy/2}}3 (e^{3iy/2}-\cos\frac{\sqrt{3}x}{2} +\sqrt{3}\sin\frac{\sqrt{3}x}{2}),\\
    j_2(x+iy,x-iy;3)&=\frac{e^{-iy/2}}3 (e^{3iy/2}-\cos\frac{\sqrt{3}x}{2} -\sqrt{3}\sin\frac{\sqrt{3}x}{2}),\\
    j_0(x+iy,x-iy;4)&=\frac12(\cos x+\cos y),\\
    j_1(x+iy,x-iy;4)&=\frac12(\sin x+i\sin y),\\
    j_2(x+iy,x-iy;4)&=\frac12(-\cos x+\cos y),\\
    j_3(x+iy,x-iy;4)&=\frac12(-\sin x+i\sin y),\\
    j_0(x+iy,x-iy;6)&=\frac13(2\cos\frac{\sqrt{3}x}{2}\cos\frac{y}{2} +\cos y),\\
    j_1(x+iy,x-iy;6)&=\frac13\big[\sqrt{3}\sin\frac{\sqrt{3}x}{2}\cos\frac{y}{2} +i(\cos\frac{\sqrt{3}x}{2}\sin\frac{y}{2}+\sin y)\big],\\
    j_2(x+iy,x-iy;6)&=\frac13\big(-\cos\frac{\sqrt{3}x}{2}\cos\frac{y}{2}+\cos y +i\sqrt{3}\sin\frac{\sqrt{3}x}{2}\sin\frac{y}{2}\big),\\
    j_3(x+iy,x-iy;6)&=-\frac{2i}3 \big(\cos\frac{\sqrt{3}x}{2}-\cos\frac{y}{2}\big) \sin\frac{y}{2}.
\end{split}
\end{equation}
The rest of $j_l(p,q;6)$ are obtained from the functions given above by means of the relations \eqref{discr_Bess_prop_a} and \eqref{discr_Bess_prop_g}. Notice that the variables $x,y\in \mathbb{C}$ in formulas \eqref{discr_Bess_func_expl}. The plots of several discrete Bessel functions are presented in Fig. \ref{discrete_Bessel_plots}.

The functions $j_l(p,q;N)$ with $N=\{1,2,3,4,6\}$ possess the translational symmetry additionally to the symmetry under rotations described in  \eqref{discr_Bess_prop_h}. The distinguished values $N=\{3,4,6\}$ are exactly the numbers of sides of the regular polygons that can pave a plane. For $N=3$, there is the translation vector $a=4\pi/\sqrt{3}$ in the complex plane. The functions $j_l(p,q;3)$ are multiplied by the phase factor $e^{-2\pi i/3}$ under the translation by $a=4\pi i/3$. Moreover, the following relation is fulfilled
\begin{equation}
    j_{\pm1}(x+iy,x-iy;3)=j_0\big(x+iy \mp\frac{4\pi}{3\sqrt{3}},x-iy \mp\frac{4\pi}{3\sqrt{3}};3\big).
\end{equation}
For $N=4$, the functions $j_l(p,q;4)$ change their sign under the translation by the vector $a=\pi(1+i)$. Besides, the relation,
\begin{equation}
    j_l(x+iy+\pi,x-iy+\pi;4)=j_{l+2}(x+iy,x-iy;4),
\end{equation}
is valid. For $N=6$, the functions $j_l(p,q;6)$ are invariant under the translation by the vector $a=4\pi\sqrt{3}$. Of course, in all the abovementioned cases one can combine the translations with the rotations by an angle of $2\pi/N$ and add the resulting vectors taken with integer coefficients.

\section{Vandermonde matrix}\label{Vandermon_Matrix}

In this appendix, we give some necessary facts about the Vandermonde matrix \cite{DemKoev06,TucWhit14,BDGY21,LiLiao21,BagMitr99,Marvasti21}. The Vandermonde matrix $V$ is defined as follows
\begin{equation}\label{Vanderm_defn}
    V_{nk}:=z_n^{k-1}, \qquad k=\overline{1,N},
\end{equation}
where $z_n$, $n=\overline{1,M}$, is some set of complex numbers. Henceforth we consider the case $M=N$. Then
\begin{equation}
    \det V=\prod_{1\leqslant i<j\leqslant N}(z_j-z_i).
\end{equation}
It is clear that the matrix $V$ is invertible when all the numbers $z_n$ are different. In that case,
\begin{equation}\label{V-1_expl}
    V^{-1}_{nk}=(-1)^{N-n}\frac{\s_{N-n}(z_1,\cdots,\hat{z}_k,\cdots,z_N)}{\prod_{m=1, m\neq k}^N(z_k-z_m)},
\end{equation}
where $\s_k(z_1,\cdots,z_l)$, $k\leqslant l$, is an elementary symmetric polynomial of $k$-th power
\begin{equation}
    \s_k(z_1,\cdots,z_l):=\sum_{1\leqslant n_1<\cdots<n_k\leqslant l}z_{n_1}\cdots z_{n_k},
\end{equation}
and the hat over $z_k$ in \eqref{V-1_expl} means that this variable is excluded.

Consider the particular case of the Vandermonde matrix \eqref{Vanderm_defn} where $z_n=w^n$ and $w=\exp(ix)$, $x\in \mathbb{R}$. Such a matrix is related to the matrix $H$ introduced in \eqref{M_matr} as
\begin{equation}\label{M_V_rel}
    H=VU,
\end{equation}
where the unitary matrix
\begin{equation}
    U_{nk}=w^{-([(N-1)/2]+1)(n-1)}\de_{nk}.
\end{equation}
If $w^n$, $n=\overline{1,N}$, are distinct, then the matrix $V$ and, consequently, the matrix $H$ are invertible. As long as
\begin{equation}
    H^\dag H= U^\dag V^\dag V U,
\end{equation}
the singular numbers of the matrices $H$ and $V$ coincide. In particular, the condition numbers of the matrices $H$ and $V$ are the same
\begin{equation}
    \kappa=\|H\|\|H^{-1}\|=\|V\|\|V^{-1}\|=s_1/s_N,
\end{equation}
where $s_k$, $k=\overline{1,N}$, are the singular numbers of the matrix $V$ numbered in descending order.

It turns out that for $|x|\ll1/N$ the following asymptotics takes place
\begin{equation}\label{sing_num}
    s_k\approx\frac{\Ga(N+k)}{\Ga(N+1-k)}\frac{\Ga^3(k)}{\Ga(2k)\Ga(2k-1)}|x|^{k-1},\qquad k=\overline{1,N}.
\end{equation}
Then the condition number is written as
\begin{equation}\label{cond_num}
    \kappa\approx\frac{N\Ga(2N-1)}{\Ga^3(N)}|x|^{1-N}\approx\frac{e}{\pi\sqrt{2}}\Big(\frac{N|x|}{4e}\Big)^{1-N},
\end{equation}
where it has been assumed $N\gg1$ in the latter approximate equality. Since $x$ is small, the condition number rapidly grows with increasing $N$. In the domain $N\gtrsim1/ |x|$, where the expression on the right-hand side becomes decreasing, formula \eqref{cond_num} is not applicable.


\end{document}